\documentclass[preprint2]{aastex} 

\begin{document} 
\title{A PERTURBED TRI-POLYTROPIC MODEL OF THE SUN}
\author{Pinz\'on G.\footnote{e-mail \textit{ gpinzon@ciencias.unal.edu.co} }, Calvo-Mozo B.\footnote{e-mail \textit{ bcalvo@ciencias.unal.edu.co} } }  
\affil{Observatorio Astron\'omico Nacional, Universidad Nacional de Colombia\\
A.A. 2584, Bogot\'a 1, Colombia\\ } 
\begin{abstract}  
\noindent \footnotesize{Based on the Solar Standard Model SSM of Bahcall and Pinsonneault (SSM-BP2000) we developed a solar model in hydrostatic equilibrium using three polytropes, each one associated to the \textit{nuclear}, the \textit{radiative} and the \textit{convective} regions of the solar interior. Then, we apply small periodic and adiabatic perturbations on this tri-polytropic model in order to obtain proper frequencies and proper functions which are in the \textbf{p-modes} range of low order  $0<l<5$; for $l=2,3$ and $4$ these values agrees with \textbf{GOLF} observational data within a few percent.}
\end{abstract} 
\keywords{Solar Standard Model (SSM), Lane-Emden, Non Radial Oscillations (NRO), p-modes}
\section{Introduction} 

\noindent Polytropic models have largely been used in the study of NRO of a gaseous sphere [Cowling,1941; Kopal,1949; Scuflaire,1974; Tassoul,1980]. 

\noindent We have computed the first modes of a tri-polytropic model (\textbf{TPM}) whose indices $n_{1}=1.50$, $n_{2}=3.78$ and $n_{3}=20$ describes the \textit{convective}, \textit{radiative} and \textit{nuclear} zones respectively of the solar interior. We used Cowling's approximation [Cowling,1941] which reduces the order of the system of differential equations to 2 (instead of 4). The radial part of the perturbation obeys equations (1) and (2) [Ledoux-Walraven,1958]:

\begin{eqnarray}
\frac{dv}{dr} &=&\left[ \frac{L_{l}^{2}}{\sigma ^{2}}-1\right] \frac{P^{%
\frac{2}{_{\Gamma _{1}}}}}{\rho }w  \label{ledoux} \\
\frac{dw}{dr} &=&\frac{1}{r^{2}}\left[ \sigma ^{2}-N^{2}\right] \frac{\rho }{%
P^{\frac{2}{\Gamma _{1}}}}v  \label{ledouxa}
\end{eqnarray}
where
\begin{equation}
v=r^{2}\delta rP^{\frac{1}{\Gamma _{1}}}
\end{equation}

\begin{equation}
w=\frac{P'}{P^{\frac{1}{\Gamma _{1}}}}
\end{equation}

\noindent are proper functions, $l$ is the degree of the spherical harmonic, $\Gamma _{1}$ is 
the adiabatic exponentent equal to  $\frac{5}{3}$, $\sigma$ is the angular
 frequency, $N$ is the \textit{Brunt-V\"{a}is\"{a}l\"{a}} frequency and  
$L_{l}$ is the \textit{Lamb} frequency.

\noindent The equations (1) and (2) and the boundary conditions
lead to a eigenvalue problem with eigenvalue $\sigma^{2}$, which is the problem to be solved.

\section{Polytropes} 
\noindent Our unperturbed model consist of a gas of particles with spherical symmetry, selfgravitating, in hydrostatic equilibrium and  with its state equation given by:

\begin{equation}
P=K\rho ^{\gamma }=K\rho ^{1+\frac{1}{n}}
\end{equation}

\noindent $K$ and $\gamma$ are parameters that depend only on the polytropic index $n$, and the mass and radius of the configuration. The  polytrope theory
developed by the ends of the XIX century, can be used to know the dynamical structure 
of a star, within which local quasistatic thermodynamic changes follows a \textit{polytropic process}, i.e. one in which the specific heat remains constant.  This approach can be used in some regions of Sun's interior. We have used the pressure and the density data from the sophisticated SSM of Bahcall and Pinsonneault to plot  $\gamma=\frac{dLnP}{dLn\rho}$ $vs$ $x$ with $x=r/R_{\odot}$. Various regions clearly emerge. Of these regions, the outer\-most one ($\gamma_{1}=1.677$, i.e. $n_{1}=1.50$) represents the \textit{convective} zone where heat transport is achieved by adiabatic convection. It's SSM output in \textit{Fig.1} is approximated rather well by a constant straight line indicating a polytropic behavior. The se\-cond intermediate zone labeled \textit{radiative zone} in \textit{Fig.1} can be approach by a polytrope $\gamma_{2}=1.264$, i.e. $n_{2}=3.788$. Finally, we have taken an ave\-rage of $\gamma$ for the innermost regions labeled  \textit{nuclear zone}, so we represent this zone by a polytrope with $\gamma_{3}=1.050$, i.e $n_{3}=20.0$. This three polytropes have been use by Hendry (1993).  

\begin{center}
\begin{figure}[h]
\begin{center}
\includegraphics[angle=0,width=7cm]{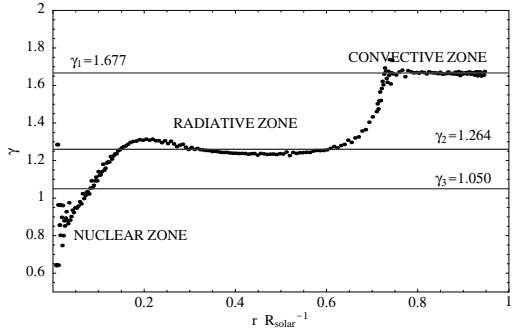}
\caption{\footnotesize {$\gamma$ at the solar interior due to SSM [Bahcall and Pinsonneault,2000]. Regions with  $\gamma$ constant are described by polytropic process. The tri-polytropic  model is obtained if one assume three values for $\gamma$; $\frac{5}{3}$ in the \textit{convective} zone, $1.264$ in the \textit{radiative} zone and $1.050$ in the \textit{nuclear} zone.}}
\end{center}
\end{figure}
\end{center}
 
\subsection{Three polytropes within the  Sun} \label{bozomath} 
\noindent Following Hendry [Hendry, 1993] we use $\xi$, $\theta$ as the variables in the \textbf{Lane-Emden} equation for 
the \textit{convective} zone with index $n_{1}$; $\eta$, $\phi$ as the variables for the \textit{radiative} zone with index $n_{2}$ and $\zeta$, $\psi$ as the variables for the \textit{nuclear} zone with index $n_{3}$. Hence the  parametric
 polytropes are:

\begin{eqnarray}
P &=&K_{1}\rho ^{\frac{5}{3}},\textrm{ with }\rho =\lambda _{1}\theta (\xi
)^{1.5}  \label{fit} \\
P &=&K_{2}\rho ^{1.26},\textrm{ with }\rho =\lambda _{2}\phi (\eta )^{3.85}\\
P &=&K_{3}\rho ^{1.05},\textrm{ with }\rho =\lambda _{3}\psi (\zeta )^{20}
\end{eqnarray}

\noindent The main challenge is to learn how to fit these three polytropes together. Since the physical quantities $P$, $\rho$ and $M$ are continuous across the interfaces (not for example $\theta$ or $\phi$), the variables $U$ and $V$ given by

\begin{equation}
U =\frac{\xi \theta ^{n}}{-\theta'} 
\end{equation}
\begin{equation}
V =\frac{(n+1)\xi (-\theta')}{\theta}
\end{equation}
\noindent are be very useful [Chandrasekhar,1939].

\begin{center}
\begin{figure}[h]
\begin{center}
\includegraphics[angle=0,width=8cm]{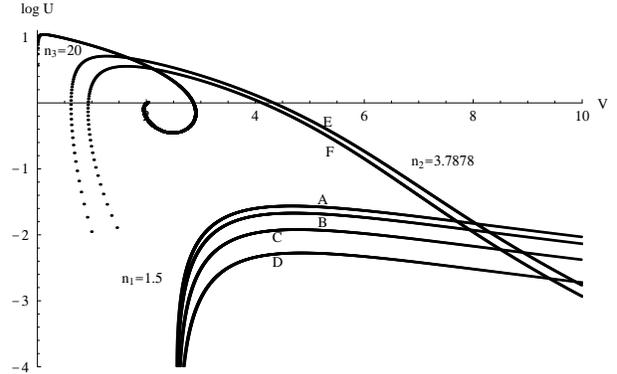}
\caption{\footnotesize {Solutions of the  \textbf{Lane-Emden} equation. The center of the Sun as at $V=0$, $U=3$ and its surface at $U=0$, $V=\infty$. 
We choose $\theta'_{\xi_{1}}=-0.00325$ (curve $C$) as a \textbf{M}-Solution for the convective zone and $\phi'_{\eta_{1}}=-0.0117$  (curve $E$) as a \textbf{M}-Solution for the radiative zone.}}
\end{center}
\end{figure}
\end{center}

\noindent Let us start by considering the \textit{convective} zone. We take $n_{1}=1.5$ thus $\xi_{1}=3.6538$ and $\theta'_{\xi_{1}}=-0.2033$. Though, this polytrope is not being used in the vecinity of $\xi=0$, it is possible to consider all of the solutions of the \textbf{Lane-Emden} equation for $n_{1}=1.50$. These may be generated beginning at $\xi_{1}$ with an arbitrary starting slope and integrating inwards. Solutions with starting slopes less negative than $\theta'_{\xi_{1}}$ are of particular interest (in the lite\-rature, they are referred to as \textbf{M}-solutions) since these are the ones which intersect the polytrope that represents the \textit{radiative} zone  $n_{2}=3.788$. Four such solutions, translated into the $U$, $V$ variables, are shown in \textit{Fig.2}, as the curves $A$, $B$, $C$ and $D$ there. Solutions with starting slopes more negative than $\theta'_{\xi_{1}}$ (\textbf{F}-solutions) do not intersect the \textit{radiative} polytrope and so do not need to be considered here. In the same way we have worked the intersection between the \textit{nuclear} and \textit{radiative} zones; we have fixed the \textit{nuclear} polytrope with the $\theta_{20}$ \textbf{E}-solution which intersects the \textbf{M}-solutions (\textit{Fig.2}, as curves $E$ and $F$) that represents the  \textit{radiative} zone  $n_{2}=3.788$, this occur at the neighbors to $0.25R_{\odot}$.

\noindent Knowing $\theta(\xi)$, $\phi(\eta)$ and $\psi(\zeta)$ we can deduce the density and pressure curves $\rho(r)$ and $P(r)$. The above model yields a central density\footnote{\textbf{SSM} value: $1.524\times10^{5}$ $Kg m^{-3}$ [Bahcall and Pinsonneault,2000]} of $\rho_{c}=1.483\times10^{5}$ $Kg m^{-3}$ in comparation with $\rho_{c}=1.299\times10^{5}$ $Kg m^{-3}$ obtained for the bi-polytropic model \textbf{BPM} [Pinz\'on-Calvo-Mozo,2001].

\noindent A comparison of our \textbf{BPM} and \textbf{TPM} models with the \textbf{SSM-BP2000} are shown in \textit{Fig.3}, showing a better fit of the \textbf{TPM} with respect to the \textbf{SSM-BP2000} in the most central part.

\begin{center}
\begin{figure}[h]
\begin{center}
\includegraphics[angle=0,width=7.5cm]{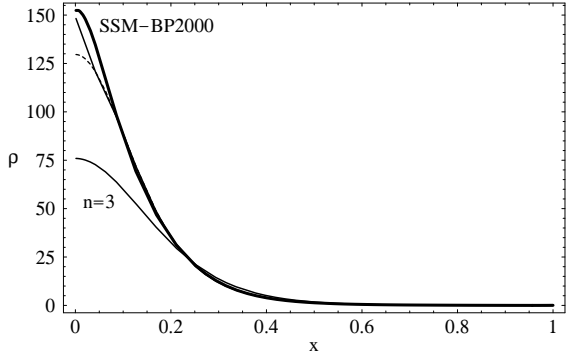}
\end{center}
\caption{\footnotesize {Density (in $g$ $\times$ $cm^{-3}$) at the solar interior for \textbf{TPM} (thin line), for the \textbf{BPM} (dashed line), for the \textbf{SSM-BP2000} (thick line) and for the $n=3$ one polytrope model.}}
\end{figure}
\end{center}

\subsection{Characteristic Frequencies}

\noindent The characteristic $N$ and $L$ frequencies so called \textit{Brunt-V\"{a}is\"{a}l\"{a}}  and \textit{Lamb} frequencies associated with restoring forces of pressure and gravity in a polytrope of order $n$ can be calculated as functions of radius in terms of the Lane-Emden function $\theta$. Thus, if we define:

\begin{equation}
\omega _{g}=\frac{GM_{\odot }}{R_{\odot }^{3}}  \label{norm}
\end{equation}

\noindent where $R_{\odot }$ and $M_{\odot }$ are the sun mass and  the sun radius, and $G$
is the gravitational constant, we can write a normalized \textit{Brunt-V\"{a}is\"{a}l\"{a}} frequency as:

\begin{equation}
N_{n}^{2}\equiv \frac{N^{2}}{\omega _{g}}=\frac{(n-n_{0})}{(n_{0}+1)}\times
CC\times \frac{3}{\theta }(\frac{d\theta }{d\xi })^{2}  \label{ncuadnorm}
\end{equation}

\noindent where  $CC$ is the central condensation  of the polytrope, defined
as the ratio of central density to mean density. The parameter $n_{0}$ is 
the effective polytropic index associated to the pulsations [Mullan-Ulrich,1988], which in turns related to the adiabatic exponent\footnote{We supposed  $\Gamma =\frac{5}{3}.$} $\Gamma$ by:

\begin{equation}
\Gamma =1+\frac{1}{n_{0}}
\end{equation}

\noindent Similarly, the value of the \textit{Lamb} frequency associated to the mode $l$ is given by:

\begin{equation}
L_{ln}\equiv \frac{L_{l}^{2}}{\omega _{g}}=3\times CC\times l(l+1)\times 
\frac{\theta }{\xi ^{2}}\times \frac{n_{0}+1}{n_{0}(n+1)}  \label{lambnorm}
\end{equation}

\noindent where $\xi $ is the Lane-Emden coordinate.

\section{NRO in a Tri-polytropic Model (TPM)}

\noindent Space oscillation properties of the solutions of equations (1) and (2) are related to the signs of the 
coefficients given by the second members of these equations. Space oscillations are allowed only in the regions
where these coefficients have opposite signs. The limits of these regions are defined by

\begin{equation}
\sigma ^{2}=\frac{l(l+1)\Gamma _{1}P}{\rho r^{2}}=L_{l}^{2}
\label{l}
\end{equation}

\begin{equation}
\sigma ^{2}=N^{2}  \label{n}
\end{equation}

\begin{center}
\begin{figure}[h]
\begin{center}
\includegraphics[angle=0,width=7.9cm]{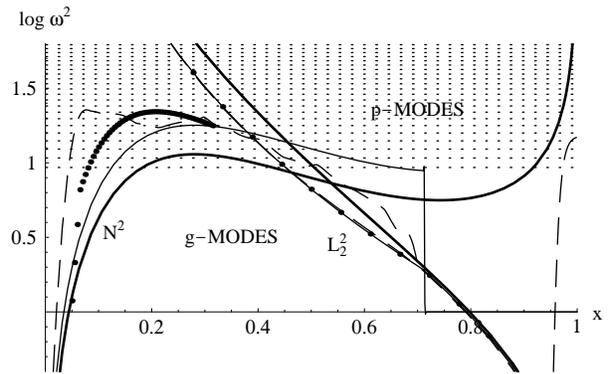}
\caption{\footnotesize {\textbf{Propagation Diagram $(l=2)$} for the \textbf{BPM} (thin lines), for the polytropic index $3$ model (thick lines) and for the 
\textbf{SSM-BP2000} (dashed lines). The \textbf{TPM} agree exactly to the \textbf{BPM} except near the center; in this region some points for the \textbf{TPM} are shown.}}
\end{center}
\end{figure} 
\end{center}

\noindent In the $(x,\omega^{2})$ plane, these equations define two curves. 
In \textit{Fig.4} we have plotted $(15)$ and $(16)$ as thin black curves for the 
\textbf{BPM}. We can see that the \textit{Lamb} frequency diverges near to the center; conversely, the \textit{Brunt-V\"{a}is\"{a}l\"{a}} runs from zero at the origin and diverges near the surface. This feature is common in all models. The \textbf{TPM} propa\-gation diagram is very similar to \textbf{BPM} except at the  nuclear region, where there is a prominent change. We have been denoted \textit{p-modes} and \textit{g-modes} the regions of $(x,\omega^{2})$  plane corresponding to the conditions of position and frequency  in the star, allo\-wing spatial oscillations. These regions are cha\-rac\-teri\-zed by the possibility of existence of progressive acoustic waves and progressive gra\-vity waves respectively  
[Scuflaire,1974]. Thus we shall refer to these regions as the acoustic and the gravity regions. We have also plotted in the same figure the frequencies (horizontal dot lines) for the first \textit{p-modes} from $p12-mode$ for the \textbf{TPM}. In addition in the same figure, we can see the propagation diagram for the \textbf{SSM} [Bahcall and Pinsonneault,2000].

\begin{center}
\begin{figure}[h]
\begin{center}
\includegraphics[angle=0,width=7.5cm]{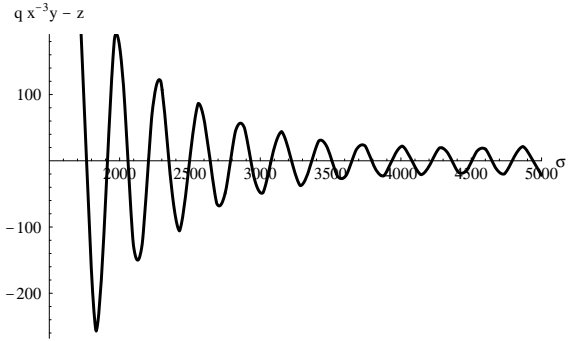}
\end{center}
\caption{\footnotesize {Behavior of the boundary condition at surface (equation $(25)$) for \textbf{TPM} model as a function of $\sigma$ in $\mu$Hz. }}
\end{figure}
\end{center}

\noindent The equations (1) and (2) are very convenient for the analytical discussion, but for the numerical computations we use the more appropriate form:

\begin{eqnarray}
\frac{dy}{dz}=\frac{l+1}{x}\left[ -y+\frac{l}{\omega ^{2}}z\right] + \frac{x}{%
\Gamma _{1}}\frac{GM_{\odot }\rho }{R_{\odot}P}\left( \frac{q}{x^{3}}y-z\right)  \label{sucuy}
\end{eqnarray}

\begin{eqnarray}
\frac{dz}{dx}=\frac{1}{x}\left[ \omega ^{2}y-lz\right] +{} R_{\odot}A\left( \frac{q}{%
x^{3}}y-z\right)  \label{sucuz}
\end{eqnarray}

\noindent where we have put:

\begin{eqnarray}
\frac{r}{R_{\odot}} &=&x \\
\frac{m}{M_{\odot }} &=&q \\
\frac{\delta(r)}{R_{\odot}} &=&x^{l-1}y \\
\frac{R_{\odot}P\prime }{GM_{\odot }\rho } &=&x^{l}z \\
\frac{R_{\odot}^{3}\sigma ^{2}}{GM_{\odot }} &=&\omega ^{2}
\end{eqnarray}

\noindent The regularity condition at the centre, first requires that  

\begin{equation}
\omega ^{2}y-lz=0  \label{center}
\end{equation}

\noindent and second, the cancellation at the surface of the lagrangian perturbation of the pressure which can be written down as:

\begin{equation}
\frac{q}{x^{3}}y-z=0  \label{surface}
\end{equation}

\noindent In order to determine the  solution uniquely, we impose the normalizing condition 
\begin{equation}
y=1 
\end{equation}

\begin{center}
\begin{figure}[h]
\begin{center}
\includegraphics[angle=0,width=8cm]{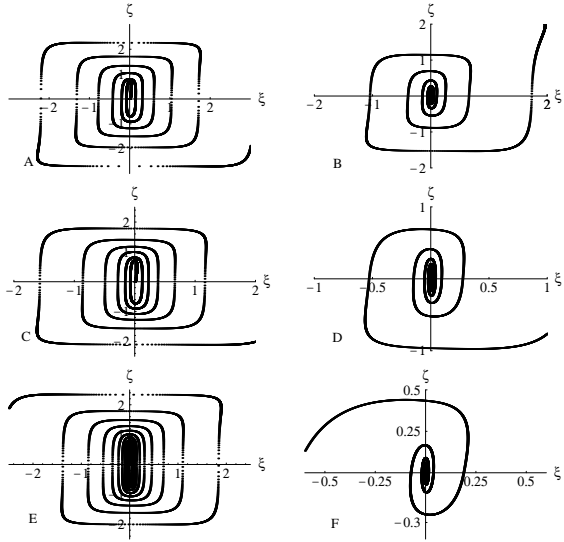}
\caption{{\footnotesize {\textbf{Some \textit{p-modes} for polytropic
 models.} Figures \textbf{A} and \textbf{B} corresponds to $p_{12}$ modes with 
$\sigma_{n,l}=\sigma_{12,2}=1796.89$ $\mu$$Hz$ and $\sigma_{12,4}=1880.45$ $\mu$$Hz$ respectively for the $n=3$ one polytrope model. 
\textbf{C} ($\sigma_{12,2}=2059.86$ $\mu$$Hz$), and \textbf{D}($\sigma_{12,4}=2165.79$ $\mu$$Hz$), are the same modes but in \textbf{BPM}, while \textbf{E} ($\sigma_{21,2}=3368.89$ $\mu$$Hz$) and  \textbf{F} ($\sigma_{21,5}=3531.43$ $\mu$$Hz$) represents the $p_{21}$ mode for \textbf{TPM}.}}}
\end{center}
\end{figure}
\end{center}

\noindent at the centre. With a trial value for $\omega ^{2}$ we integrate equations (17) and (18), with initial conditions (24) and (26) using Runge-Kutta method, with a  step size taken from a paper of Christensen-Dalsgaard, equation 
$(A.55)$ [Christensen-Dalsgaard and Mullan,1994]. Usually this solution does not satisfy equation (25) and a new integration is performed with a\-no\-ther value of  $\omega ^{2}$. This procedure is repeated until equation (25) is satisfied, using a Newton-Raphson method to improve the value of  $\omega ^{2}$.

\begin{center}
\begin{figure}[h]
\begin{center}
\includegraphics[angle=0,width=7.5cm]{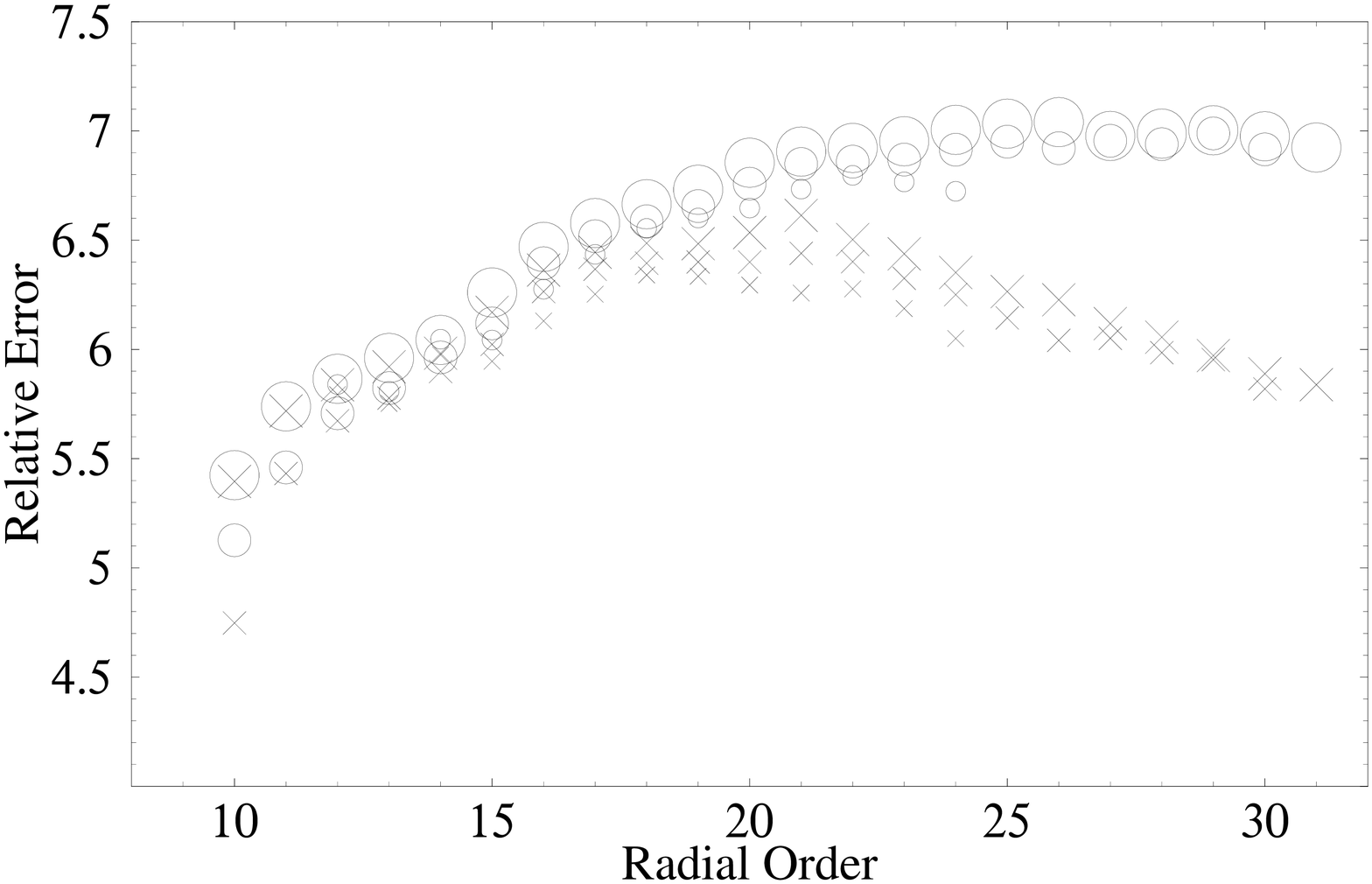}
\end{center}
\caption{\footnotesize {Comparison of \textbf{GOLF} data with those of the \textbf{TPM} (crosses) and the \textbf{BPM} (open circles) models for $l=2,3,4$; sizes of circles and crosses are inversely proportional to $l$. The \textit{Relative Error} value is given by \footnotesize{$\frac{\sigma_{MODEL}-\sigma_{GOLF}}{\sigma_{GOLF}}\times100\%$}}}
\end{figure}
\end{center}

\subsection{Phase Diagram}

\noindent The radial displacement  $\delta(r)$ and the pressure perturbation 
$P'$ are periodic space functions; the variables $v(r)$ and $w(r)$ vary strongly from
the center to the surface, then it is impossible to plot them directly along the axes. However, the most appro\-piate functions:

\begin{equation}
\xi =\pm \log _{10}(1+\left| \frac{\delta(r)}{R_{\odot}}\right|)
\end{equation}

\begin{equation}
\zeta =\pm \log _{10}(1+\left| \frac{R_{\odot}P'}{GM_{\odot }\rho }\right|)
\end{equation}

\noindent have been plotted $(27)$ and $(28)$ in each axis of the \textit{Fig.6}; their signs are chosen according to the signs of the variables $\delta(r)$ and $P'$. The number of intersections of each curve with the ordinate axis in \textit{Fig.6} (the origin excluded) is equal to the order of the mode. The sense of rotation of this curves is a characteristic feature for the \textbf{p-modes} [Scuflaire,1974].

\section{Conclusion}
\noindent  Although we do not use an atmosphere model and the input physics is described by a multi-polytropic structure, the modes obtained 
from the \textbf{TPM} are close to the observacional data. We show in \textit{Fig.7} a comparison  with GOLF data\footnote{taken from www.medoc.ias.u-psud.fr/golf.html}, showing an agreement within a few percent (roughly between $4.5 \%$ and $7 \%$) up to a radial order of $30$. However, we can see that the \textbf{TPM} model is more reliable than the \textbf{BPM} one for all radial orders considered ($n_{r}=10$ to $30$), being noticeable at the higher ones.

\bigskip

\end{document}